# Strong-field electro-optic sampling of light

**Anton Husakou[1], Nicholas Karpowicz[2,3], Vladislav S. Yakovlev[2,3], Misha Ivanov[1], Kazuhiro Yabana[4], AND Dmitry A. Zimin[5,*]**

[1]*Max Born Institute for Nonlinear Optics and Short Pulse Spectroscopy, Berlin, Germany.*
[2]*Max-Planck-Institut für Quantenoptik, Garching, Germany.*
[3]*Fakultüt für Physik, Ludwig-Maximilians-Universität, Garching, Germany.*
[4]*Center for Computational Sciences, University of Tsukuba, Tsukuba, Japan.*
[5]*Laboratory of Physical Chemistry, ETH Zürich, Zürich, Switzerland.*
[*]*Corresponding author email: dzimin@ethz.ch*

**Abstract:** We report the experimental realization of electro-optic sampling of light utilizing strong-field light–matter interaction. We show that in contrast to the perturbative regime where the non-linear signal (eg. sum-frequency generation) is localized at the narrowband spectrum that must match the local oscillator, the strong-field regime relies on the generation of a supercontinuum that automatically fulfills the overlaps with a local oscillator providing extremely broadband detection bandwidth. We develop a semi-classical model, benchmark it against the experiments with carrier-envelope-phase (CEP) stable sub-3 femtosecond pulses of octave-spanning bandwidth, and further validate it by a first-principle time-dependent density functional (TDDFT) theory. Our work establishes a simple electro-optic approach for investigating broadband high-frequency dynamics of matter on sub-cycle timescales with attosecond precision.

## 1. Introduction

Upon light-matter interaction, the spatio-temporal evolution of the electric field brings matter into motion. The response of matter evolves in time and space and can be imprinted on the interacting electric field. By reading out this imprint, one gains access to the underlying physical processes responses of which can span across various timescales and eigenfrequencies that serve as fingerprints of these processes [1]. In contrast to conventional spectroscopy, the electric-field resolved approach known as fieldoscopy [2], enables direct tracking of sub-cycle matter dynamics *during* the light–matter interaction. For instance it has been recently used to probe the formation of optical response from electron–hole plasma [3,4], ultrafast magnetism [5], exciton coherence [6], and light–matter energy transfer in photoexcited solids [7]. It has also been demonstrated that the electric field itself can be used to encode digital information [8], and serve as a logic gate [9] or a memory [10], paving the way toward on-chip [11] petahertz (1 PHz = $10^{15}$ Hz) optoelectronics [12]. For this, precise measurement of the spatio-temporal profile of the electric field of light is essential.

Sampling the electric field, however, remains a significant challenge and is a central focus of contemporary research [13,14]. Available techniques can be classified as gating and heterodyne electro-optic approaches. Each technique is characterized by the frequencies to which it provides access and therefore fingerprints of which physical processes can be studied; the detection bandwidth, which defines the temporal resolution; and the sensitivity that determines the strength of physical processes that can be observed. For instance, vibrational motion in matter typically occurs in the mid-infrared range and picosecond time scales; excitonic and plasmonic dynamics span the infrared to ultraviolet regions and femtosecond timescales; electronic transitions however can extend into the extreme ultraviolet (XUV) or X-ray domains [15] and attosecond or even zeptosecond timescales.

Electronic and optical gating [16] can be implemented through various mechanisms: in the Auston switch [17], by modulating the material's conductivity; in the attosecond streak camera [18], by single-photon ionization; in linear [19] and nonlinear [20,21] photoconductive sampling, by photoinjection [22]; in the TIPTOE [11,23,24], by tunneling ionization; in ARIES [25,26] by a perturbation of the high-harmonic generation. Gating exhibit drawbacks: i) the bandwidth rapidly decreases as the gate duration approaches the half-cycle of the measured frequency. For example, sampling 500 nm light already requires a gate duration in the attosecond regime, necessitating intense ultrashort pulses, highly nonlinear light–matter interaction, and complex experimental infrastructure; ii) the non-optical nature prevents using optical cameras (despite attempts towards spatial extension [27,28]); iii) the sensitivity and the dynamic range are far behind the heterodyne electro-optic methods [29–31].

Heterodyne electro-optic sampling [31–38] is the most common fieldoscopic approach due to its simplicity, all-optical nature, and enormous sensitivity down to single photons [39]. It relies on the generation of a byproduct electric field (BPF) via perturbative light-matter interaction, such as for instance sum-frequency generation, and a subsequent interference with a local oscillator (LO). This simple approach enables access to the electric field in both time [17,33–35] and space [36], offers exceptional sensitivity and dynamic range [29–31], and does not require complex infrastructure or ultrashort few-cycle pulses. Crucially, this approach does not require an optical gate to be shorter than half the period of the measured frequency [31], which is a major limitation in the gating schemes. However, the electro-optic response is either restricted to low frequencies [32–36] or exhibits a bandpass-like (non-continuous) character [31]. These limitations are fundamentally tied to the perturbative light-matter interactions that electro-optic sampling rely upon (see Supplementary Material for details). The perturbative non-linear signal (eg. sum-frequency generation) is localized at the upconverted bandwidth and must match the local oscillator. This becomes challenging for high petahertz-scale frequencies. For instance, sampling of ~ 0.5 PHz light with conventional electro-optic sampling already requires [34] an extremely short visible–UV local oscillator.

In this work we report on the alternative route. Unlike perturbative regime, we utilize highly nonlinear light-matter interaction, such as Brunel radiation [40] (arising from time-dependent plasma density), the recently observed injection current [41], which results from electron displacement during tunneling, and cascaded nonlinearities (sequential generation of new spectral components during light-matter interaction). This results in the generation of a broadband continuum that automatically fulfills the spectral overlap with a local oscillator for many different test frequencies, thereby overcoming conventional electro-optic limitations. We realize this regime experimentally and develop a physical model. We find that the strong-field regime can enable continuous multi-petahertz bandwidth, which we further validate by the first-principles time-dependent density functional theory (TDDFT).

## 2. Results

The idea of electro-optic sampling is to combine the pulse to be measured (test, with the field $E_T(t)$) with another (sampling, with the field $E_S(t-\tau)$) and to produce a byproduct electric field ($E_{BPF}(t,\tau)$) that is linearly proportional to $E_T(t)$. When $E_{BPF}(t,\tau)$ is superimposed with a local oscillator $E_{LO}(t)$ and is guided to a detector (e.g. photodiode), the signal ($S$) is proportional to the intensity of the heterodyne interference at the detection frequency $\omega_D$:

$$S(\tau) \propto \int (E_{LO}(t,\tau,\omega_D) + E_{BPF}(t,\tau,\omega_D))^2 dt. \quad (1)$$

When $E_{LO}(t)$ is strong and $E_{BPF}(t,\tau)$ is weak, the dominant $\tau$-dependent term is:

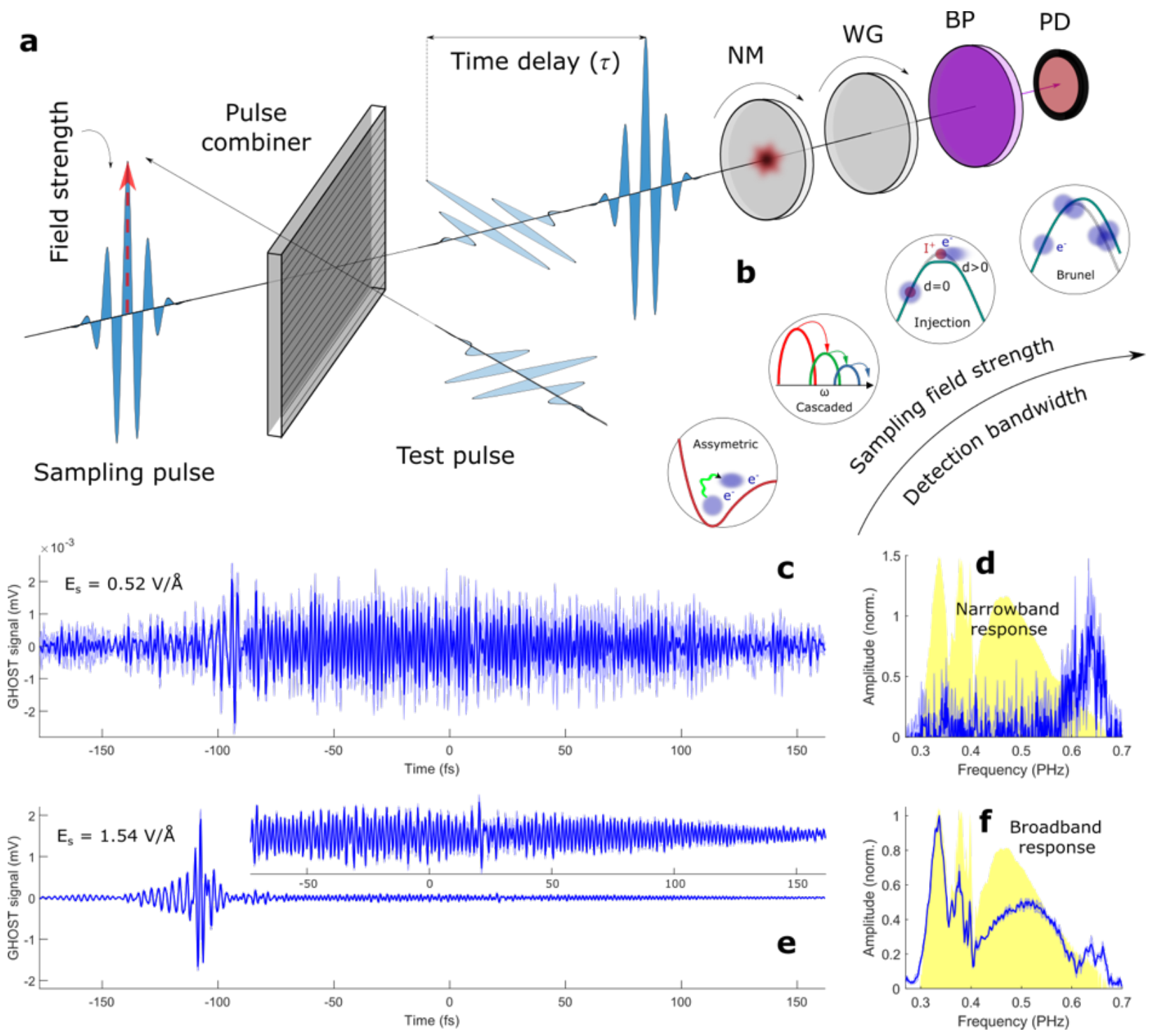

Fig. 1. **Summary of the experiment.** **a**, Schematic of the experiment in which two cross-polarized pulses, sampling and test, are recombined with a pulse combiner and focused on a nonlinear medium (NM). The test pulse is kept weak, while the field strength of the sampling pulse varies from 0.52 to 1.54 V/Å. The medium produces LO and BPF that are guided to a photodiode (PD) for the detection of the heterodyne intensity. The wire-grid polarizer (WG) is used to project cross-polarized BPF and LO onto one plane. The bandpass filter (BP) is used to select an interference frequency on the photodiode. **b**, Schematic of the strong-field concept. As the sampling field strength is increased, more nonlinear processes become accessible, which leads to a broadening of the spectral response. **c**, Measured photodiode signal at various time delays (τ) between two pulses in the weak-field regime. **e**, Measured photodiode signal as a function of time delay (τ) between two pulses in the strong-field regime. **d**, **f**, Spectrum of the measured signals in **c**, **e** obtained by a Fourier transformation. The yellow area corresponds to a test pulse spectrum measured with a grating spectrometer.

$$S(\tau) \propto \int E_{LO}^{*}(t,\tau,\omega_D) E_{BPF}(t,\tau,\omega_D)\, dt. \quad (2)$$

The relation between the measured signal $S(\tau)$ and the actual test electric field $E_T(t)$ constitutes the response function $R(\omega)$ of the measurement. When $R(\omega)$ is known for all frequencies, any test electric field $E_T(t)$ can be directly obtained from the measured $S(\tau)$:

$$E_T(t) \propto \mathcal{F}^{-1}\{S(\omega)/R(\omega)\}. \quad (3)$$

where, $(\mathcal{F}^{-1})$ stands for inverse Fourier transformation. The limitation of electro-optic sampling is bound to fulfilling the spectral overlap between the local oscillator and the byproduct electric field. Fulfilling this condition with perturbative light-matter interaction over large bandwidth is difficult, since the spectrum of the byproduct electric field is different for different test frequencies (see Supplementary Material for details). In other words, different test frequencies, require different local oscillator frequencies. Achieving a broadband detection on the other hand requires that $E_{BPF}(t)$ for different test frequencies always has a spectral component at the local oscillator frequency. As we show further, this condition can be realized with strong-field light-matter interaction (more details in the Supplementary Material).

We start with a conventional perturbative regime with a well-known spectral response and study whether this response can be changed upon enabling non-perturbative strong-field processes. For this we employ an octave-spanning (450 – 1000 nm), CEP–stable, sub-3 fs pulse [20], which is split into two arms: sampling and test. After independently setting the field strength and CEP of each arm, the pulses are recombined and co-linearly focused onto a nonlinear medium — a ~12 μm-thick z-cut α-quartz crystal. The test pulse is kept weak, while the strength of the sampling pulse is varied between 0.52 and 1.54 V/Å.

We begin by setting a moderate field strength of 0.52 V/Å. This value is sufficient to drive second- and third-order perturbative processes arising from the anharmonicity of electron motion [42] (see Fig. 1b), but too weak to initiate strong-field phenomena [7]. Under these conditions, we implement perturbative GHOST-like [31] electro-optic sampling, utilizing the second harmonic (SHG) of the sampling pulse as LO and four-wave mixing (FWM) between the sampling and test pulses as the BPF. The resulting SHG+FWM configuration exhibits a bandpass-like spectral response that covers only the edges of the test spectrum (Fig. 5a). Fig. 1c,d shows the measured $S(\tau, \omega_0)$ and a corresponding spectrum. As expected (see Supplementary Material for details), this measurement captures only a limited portion of the test bandwidth, highlighting limitations of perturbative electro-optic sampling.

We now attempt to activate the strong-field regime by increasing the sampling field strength to 1.54 V/Å. If strong-field processes would not generate sufficient byproduct fields, the response would remain largely unchanged regardless of the sampling field strength. In stark contrast to the perturbative regime, the recorded trace (Fig. 1e) now covers the entire test bandwidth (Fig. 1f). We note that this spectral response is not possible with conventional perturbative electro-optic sampling, which indicates a key role of higher-order nonlinear light-matter interactions. Equally important is a signal enhancement in sensitivity of approximately three orders of magnitude, accompanied by a high dynamic range—defined as the ratio between the strongest and weakest detected signal—of about 40 decibels. We note that the highest sampling field strength (1.54 V/Å) in our experiment is well below the damage threshold of quartz reported for similar few cycle pulses [7]. Our measurements were therefore well reproducible without noticeable cumulative or permanent modifications of the non-linear medium.

We identify three distinct mechanisms that define what we refer to as the strong-field regime. At elevated intensities, newly generated frequency components contribute to nonlinear wave mixing during propagation, giving rise to cascaded nonlinear effects (Fig. 1b). At higher field strengths, electrons are promoted to the conduction band, resulting in two additional processes: injection current [41] and Brunel radiation [40] (Fig. 1b). So far, only electronic gating techniques [20,43] have exploited these strong-field effects. In contrast, our approach utilizes them to generate coherent byproduct electric fields, thereby effectively transforming gating into electro-optic scheme.

We repeat the experiment shown in Fig. 1 for various sampling field strengths to investigate the transition from the perturbative regime with a narrowband response towards a strong-field regime with a broadband response (Fig. 2a-d). The experiment indicates that by entering the strong-field regime, the spectral response contains two distinct parts: 1) conventional perturbative response; 2) A new (broadband) response originating from strong-field mechanisms. We find that at very high field strengths close to the damage threshold, the strong-field part becomes dominant.

To understand this transition, we modeled the nonlinear propagation of light through the medium, accounting for strong-field processes (more in Methods and Supplementary Material).

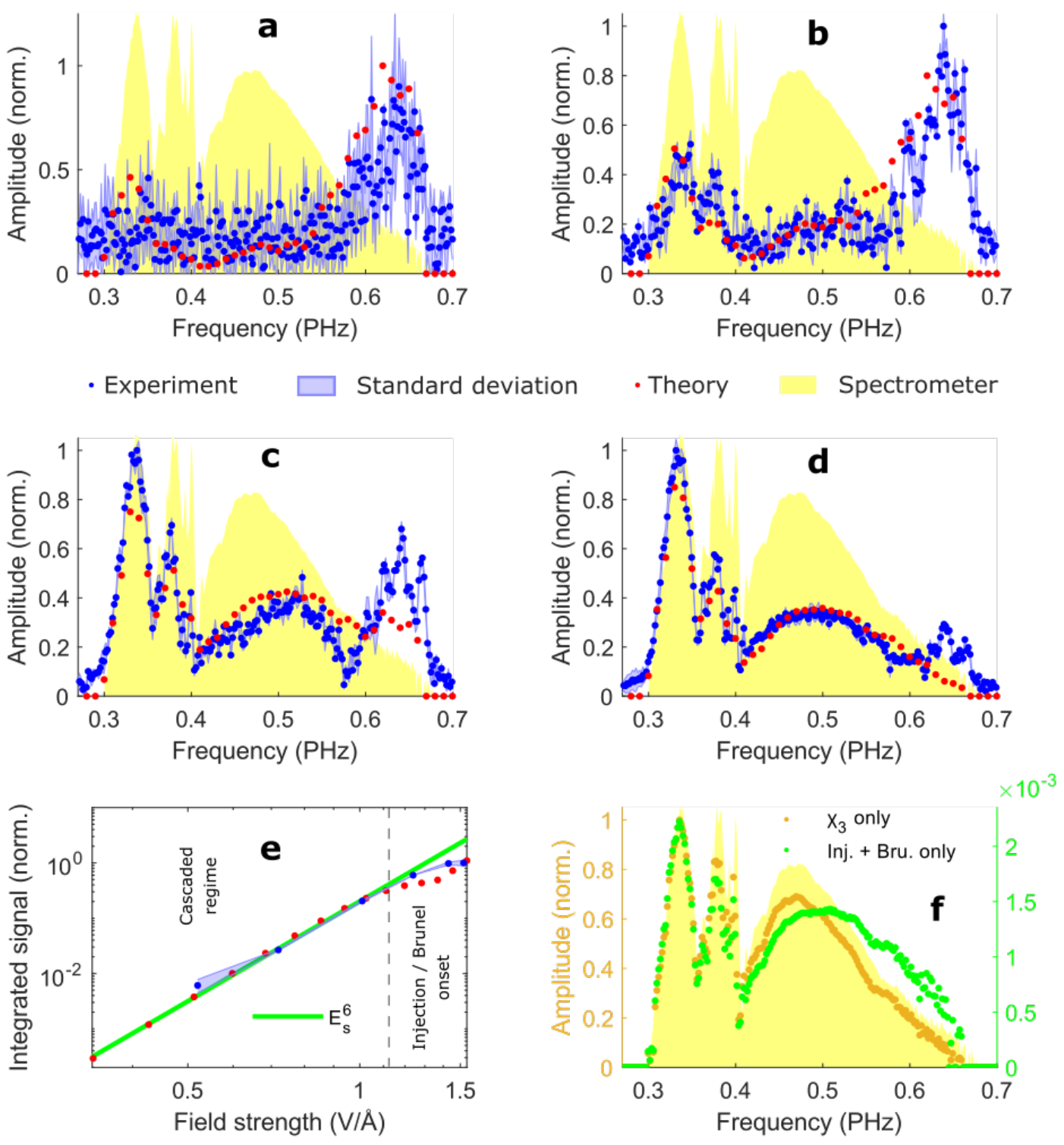

Fig. 2. **Transition from the narrowband to the broadband regime. a**, **b**, **c**, **d**, Spectra of the measured and simulated signal for the sampling pulse field strengths of 0.52, 1.01, 1.27, and 1.54 V/Å, correspondingly. The yellow shaded area indicates the spectrum of the test pulse, and the shaded blue area denotes the standard deviation. **e**, Measured and simulated scaling of the integrated signal around 0.45 PHz as a function of the field strength of the sampling pulse. **f**, The role of the third-order optical nonlinearity, injection current, and Brunel radiation in the total signal for 1.54 V/Å sampling field strength.

We employ forward-propagation equation [44] without slowly-varying envelope approximation and automatically include cascaded effects of propagation, as well as group velocity dispersion to all orders, linear absorption, and self-steepening. The good agreement of the model with *all* our experiment, suggests that it captures most relevant physics.

We now turn to unraveling the underlying physics. In the perturbative regime, the response is primarily confined around 0.65 PHz (Fig. 2a), whereas in the strong-field regime, it extends across the entire available bandwidth. To disentangle the contributing nonlinearities, we analyze the scaling behavior of the signal at 0.45 PHz and 0.65 PHz (Fig. 2e). We integrated the signal around 0.65 PHz and find that it scales with the fourth power of the sampling field strength, $I_S \propto E_S^4$. This is consistent with the SHG+FWM electro-optic sampling, which involves two sampling photons for LO and two for BPF. Interestingly, at 0.45 PHz, two distinct regimes emerge (Fig. 2e): below approximately 1.1 V/Å, the signal scales as $E_S^6$, indicating involvement of strong-field processes. Above this threshold, a deviation from the distinct $E_S^6$ scaling is observed, signaling the onset of mechanisms (Brunel radiation and injection current) that involve photoexcitation of charge carriers. Both scaling regimes are well reproduced by our model.

Fig. 2f compares the simulated contributions from cascaded nonlinear processes, Brunel radiation, and injection current dynamics, confirming that strong-field effects can be harnessed for electro-optic field sampling. The Brunel and injection current responses are highly dependent on the sampling field strength and are maximized near the material's damage threshold. Furthermore, photoinjection of charge carriers is primarily confined to the front surface of the nonlinear medium (see Supplementary Material), whereas the signal originating from perturbative light–matter interactions accumulates throughout the full thickness of the sample. As a result, Brunel and injection current contributions are most effectively generated using strong sampling fields and thin media.

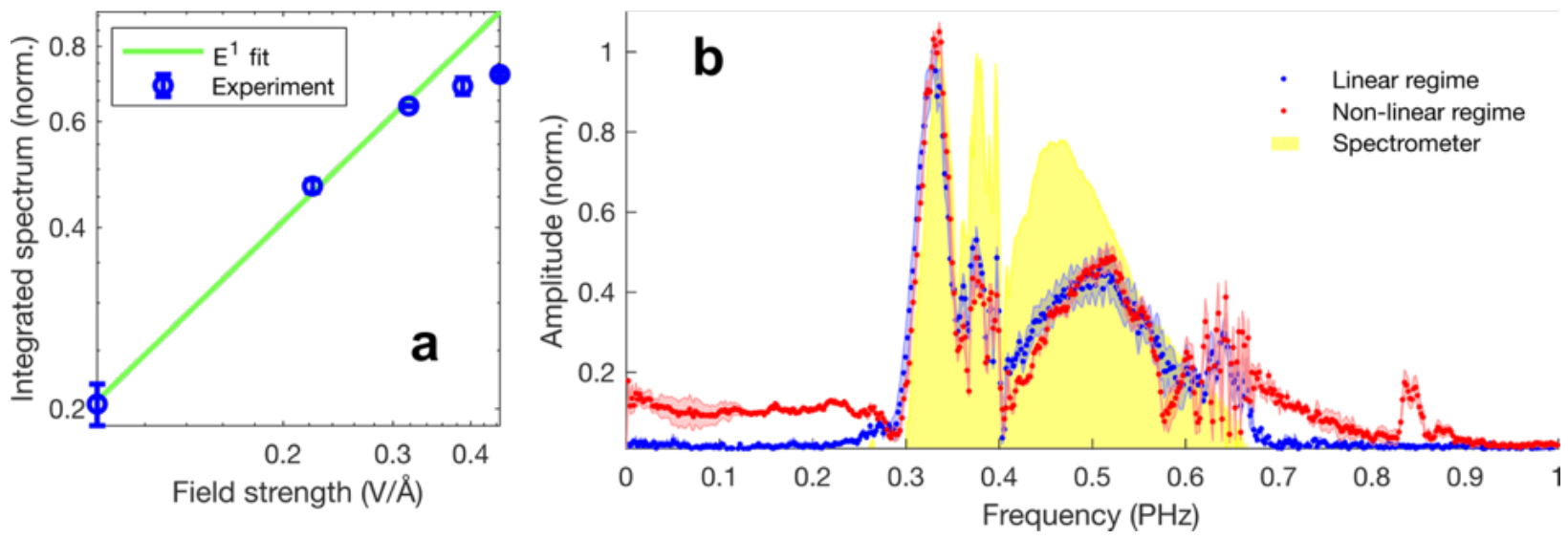

Fig. 3. **Linearity of the field sampling in the strong-field regime. a**, Measured scaling of the integrated signal around 0.45 PHz as a function of the test pulse field strength. **b**, Spectra of the measured signal in the linear and non-linear regimes.

Fieldoscopic studies rely upon a linear dependence of the measured signal on the test electric field. To validate whether this condition holds in the strong-field regime, we measure the scaling of the strong-field signal with the test field strength. Similarly to Fig. 2e, we integrate the spectrum of the measured signal around 0.45 PHz and plot its dependence on the test field strength (Fig. 3a). We find that the linearity of the strong-field regime persists up to about 0.3 V/Å (the sampling field is at least four times stronger than the test field). As we show in the Supplementary Material and further in Fig. 6e, this linearity condition is well explained by our model. In Fig. 3b we plot a comparison of the fieldoscopic measurements in the linear and non-linear regimes.

To confirm that the strong-field response dominates at high field strengths, we record fieldoscopic traces as a function of the CEP of the sampling pulse (Fig. 4d). Unlike in the perturbative regime, the spectral response in the strong-field regime is firmly CEP-dependent, exhibiting a characteristic $\pi$-periodicity. The strongest and smoothest responses are expected for CEP values of 0, $\pi$, and $2\pi$, where the sampling field features a single, well-defined electric field maximum. This behavior is confirmed both experimentally and numerically: we observe a broad spectral response for CEPs near 0 and $\pi$, and a pronounced dip at $\pi/2$ (Figs. 4c,d).

Finally, we benchmark our model against the strong-field experiment $E_S = 1.54$ V/Å) directly in the time and frequency domains (Fig. 4a,b). Quantitatively, the deviation between the model and the experiment is approximately 7 %.

## 3. Discussion

As the sampling pulse nonlinearly propagates through the medium, additional frequency components are generated. In the perturbative regime, the yield of these new components is negligibly small so that mainly the original sampling spectrum contributes to nonlinear mixing with the test photons, producing the byproduct electric field. In contrast, in the strong-field regime, the newly generated light has sufficient yield to participate in the mixing process. While it is not possible to switch off these cascaded processes experimentally, they can be selectively suppressed in our simulation (Fig. 5a,b). The observed broadening of the response and the emergence of signal components between 0.35 PHz and 0.5 PHz are clearly linked to the cascaded dynamics—without it, the spectral shape resembles that observed at low intensity. Several nonlinearities that can lead to cascaded spectral broadening are self-phase modulation, self-steepening, third harmonic generation and four-wave mixing. In the framework of the slowly varying envelope approximation, it would be easy to separate the above contributions since they correspond to different terms. However, in the case of very short pulses, such approximation cannot be used, and the nonlinear polarization $P_{NL}(t) = \epsilon_0 \chi_3 E^3(t)$ cannot be trivially split into different mechanisms. To address this, we have developed an approach (see Supplementary Material) which allows separating frequency mixing with the identical signs (IS): $\omega = +\omega_1 + \omega_2 + \omega_3$, and with different signs (DS): $\omega = +\omega_1 + \omega_2 - \omega_3$. The IS term incorporates the sum-frequency mixing of three photons, such as the mixing of two sampling photons and one test photon ($\omega = +\omega_S + \omega_s + S + \omega_T$), as well as third harmonic generation. The DS term describes several parametric and mixing processes: ($\omega = +\omega_S - \omega_S + \omega_T$, $\omega = +\omega_S + \omega_S - \omega_T$, $\omega = +\omega_S - \omega_S - \omega_T$, etc.) and includes further contributions such as self-phase modulation. In Fig. 5c,d we show the role of the IS and DS terms in the response function $R(\omega)$ for the low- and high-intensity cases. For low-intensity case, the low-frequency term consists of both IS and DS contributions while the high-frequency peak is generated solely by the DS term. For higher intensity, it is the DS term that provides spectral broadening of the response function and broadband detection.

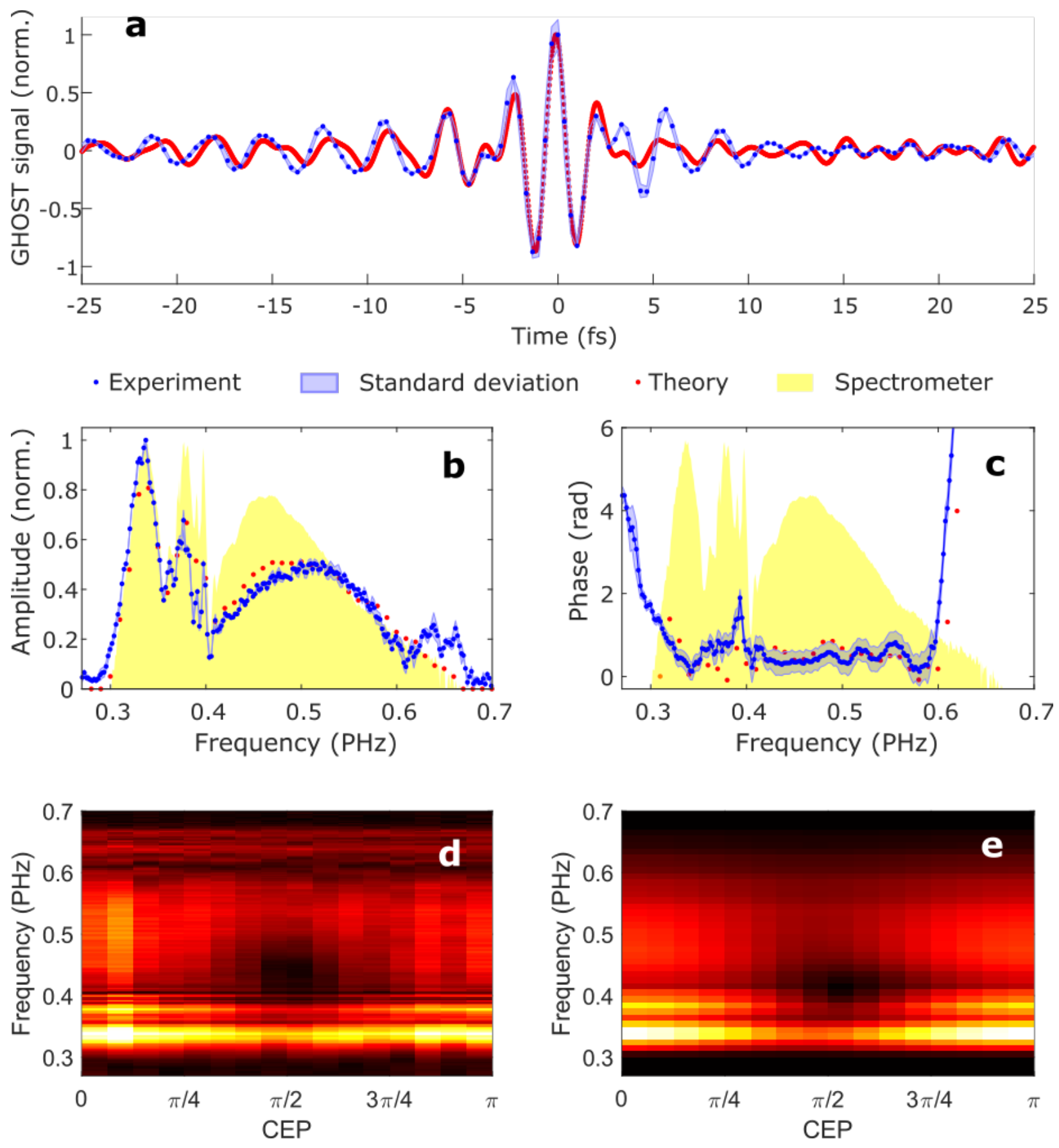

Fig. 4. **Benchmarking of the model against the experiment.** **a**, The experimental (blue dots) and numerical (red dots) temporal profiles of the signal for the sampling pulse field of 1.54 V/Å. **b** and **c**, Amplitudes and phases of the measured spectrum, correspondingly. The experimental (blue dots) and numerical (red dots) amplitudes and phases, together with the yellow shaded area which indicates the spectrum of the test pulse. The shaded blue area denotes the standard deviation. **d**, The dependence of the measured spectrum on the CEP of the sampling pulse. **e**, The simulated dependence of the spectrum on the CEP of the sampling pulse.

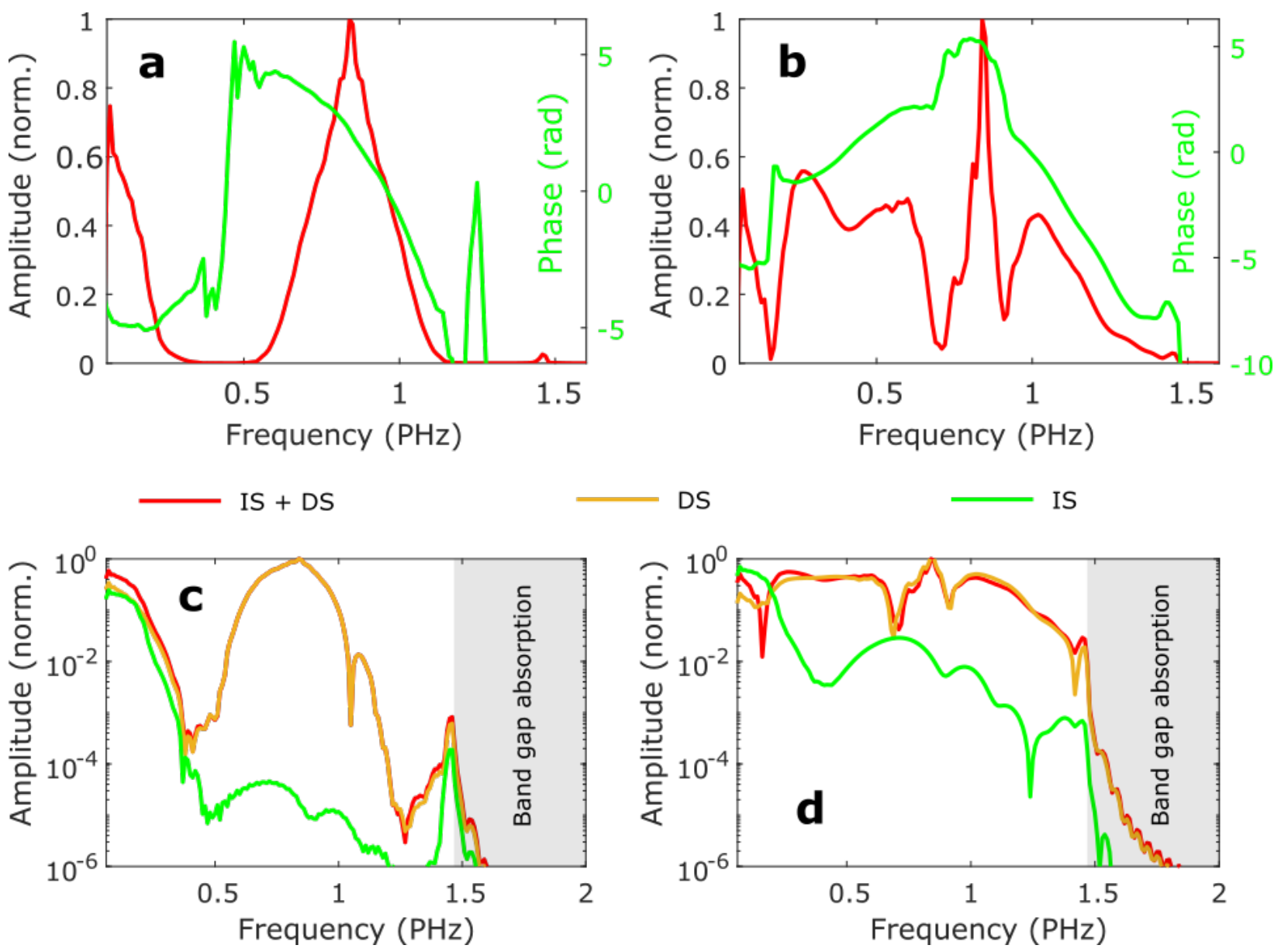

Fig. 5. **Disentangling cascaded contributions.** **a** and **b**, The amplitudes (red curves) and spectral phases (green curves) of the simulated signal with cascaded processes switched off (**a**) and on (**b**). The field strength of the sampling pulse is 1.54 V/Å. **c** and **d**, The simulated spectra with both IS and DS third-order optical processes included (red curves), only DS processes included (orange curves), and only IS processes included (green curves).

At higher sampling field strengths, the photoinjection-related mechanisms (Brunel radiation and injection current) begin to play a comparable role. In the perturbative regime, the sampling pulse is not intense enough to photoexcite electrons from the valence band to the conduction band. However, in the strong-field regime, the sampling field becomes sufficient to induce such photoexcitation. In the case of injection current, as recently demonstrated [41], the excitation of carriers from the valence to the conduction band leads to a net current and the subsequent emission of coherent radiation. The local electric field of the test pulse can perturb this excitation process, resulting in a modification of the emitted coherent light. In case of the Brunel dynamics, the charges can be subsequently accelerated by the test electric field, which in conjunction with time-dependent conduction-band electron density produce a broadband coherent emission (see Supplementary Material). The coherent light from injection current and Brunel radiation can be interfered with a local oscillator, enabling electro-optic sampling of the test electric field.

We note that the strong-field response across different spectral regions originates from the same physical mechanisms. Therefore, our broadband experiment can be used to establish a physical model by benchmarking it against the experiment. After confirming that our model reproduces all our measurements, we now simulate the overall response function (0 - 5 PHz) of the strong-field regime under realistic experimental conditions. Since the Brunel radiation and injection current are mainly confined to the surface (see Supplementary Material), it is beneficial to use thin nm-scale medium. This also minimizes absorption of photons with energies above the bandgap. Equally important, high-order nonlinearities are enabled at high field strengths, therefore we consider a $SiO_2$ (Fig. 6a,c) and diamond (Fig. 6b,d) that can sustain high intensity. These samples also provide high 3rd-order nonlinearity ($3\times10^{-20}$ $cm^2/W$ [45] and $10\times10^{-20}$

cm$^2$/W [46], correspondingly) necessary for cascaded processes, large bandgaps (9 eV [45] and 5.5 eV [47], correspondingly) and can be made as a free-standing films of nm-scale thickness. We note however that although we only consider solids in our simulation, gases or liquids can also serve as a non-linear media.

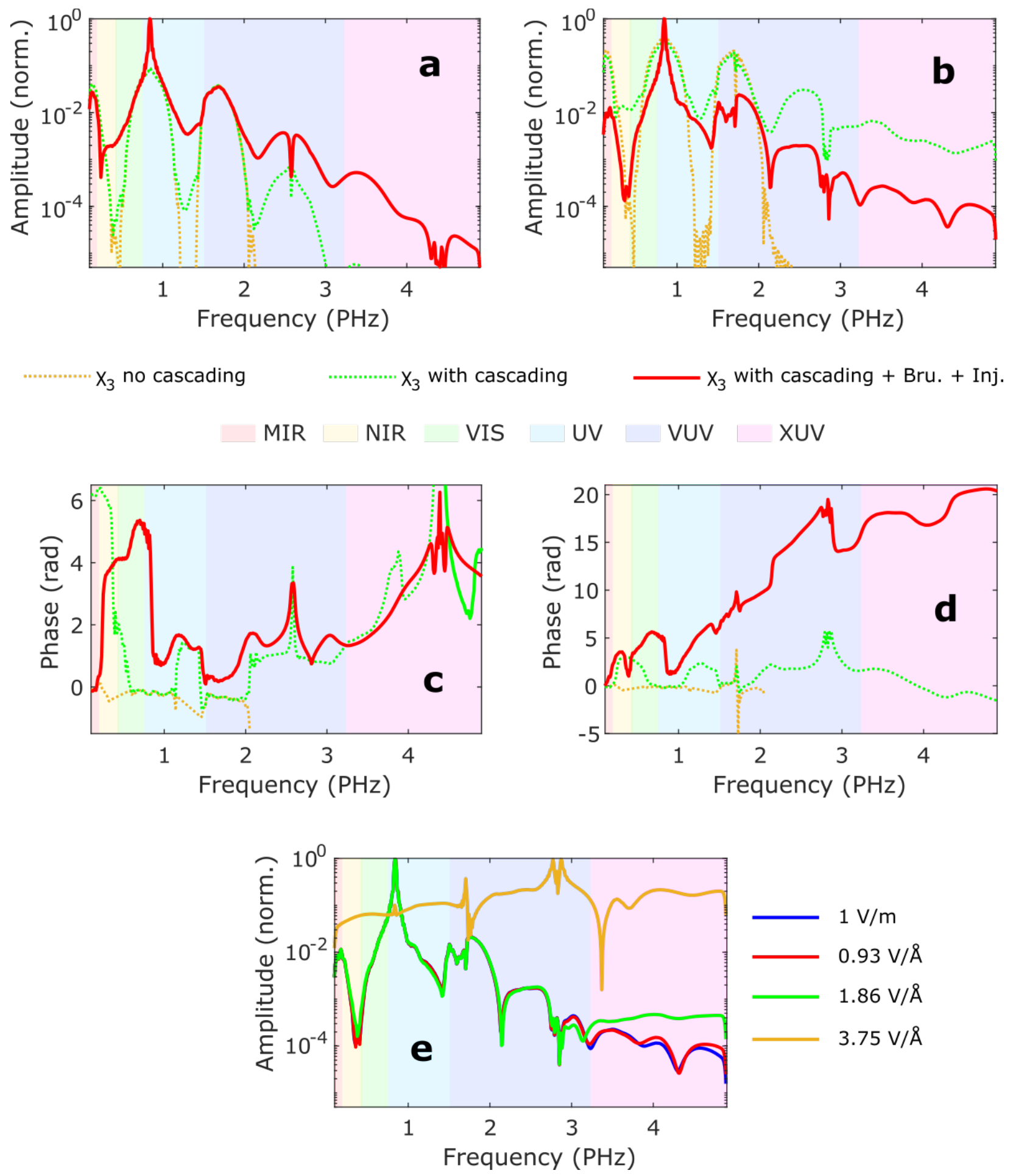

Fig. 6. **Perspectives of the strong-field electro-optic sampling. a**, The simulated response function for the thin 20-nm diamond sample with the sampling field strength of 5 V/Å. The orange curve indicates the spectrum when only third-order nonlinear processes without cascading are included, the green curve shows the spectrum when third-order nonlinear processes with cascading are included, and the red curve represents the case when third-order nonlinearity, Brunel radiation, and injection current are fully included. **b**, The simulated response function for the thin 20-nm SiO2 sample with sampling field strength of 4 V/Å. The meaning of the colors is the same as in **a**. **c** and **d**, Spectral phases of the response function for the cases shown in **a** and **b**, correspondingly. The meaning of the colors is the same as in **a**, **b**. **e**, Dependence of the normalized spectral response on the test field strength for 20-nm diamond sample and the sampling field strength of 5 V/Å.

Fig. 6 shows the full response of the strong field regime simulated for our experimental conditions. In contrast to conventional perturbative electro-optic sampling, the strong field regime is clearly much more broadband and extends up to about 5 PHz (60 nm wavelength) in both samples. As we describe in the Supplementary Material, the origin of the broadband response in the strong field regime is linked with the broadband, continuum-like emission of the byproduct electric field, which is not the case in the conventional, perturbative regime. Interestingly, we find that in diamond (Fig. 6a), Brunel radiation and injection current dominate the response, whereas in $SiO_2$ (Fig. 6b), cascaded nonlinearities are the primary contributors. This difference can be attributed to diamond's smaller bandgap, which facilitates photoexcitation and enhances the efficiency of injection current and Brunel mechanisms. When both photoionization-related and cascaded effects are numerically suppressed (orange curves in Fig. 6a,b), the simulated response reverts to that of a conventional perturbative electro-optic sampling. This further underscores the essential role of Brunel radiation and injection current in enabling a continuous and broadband response. Furthermore, Fig. 6e, shows that the spectral response remains linear with respect to the test electric field when the sampling field is at least about 5 times stronger. This is consistent with our experiment shown in Fig. 3a.

To further validate the broadband nature of the strong-field regime, we performed first-principle *ab initio* simulations based on the time-dependent density functional theory (TDDFT), using the SALMON package [48], which is known for its high accuracy. The results (see Supplementary Material) are consistent with our semiclassical model and confirm the predicted extension of the bandwidth up to 5.5 PHz.

## 4. Conclusion

We have demonstrated the strong-field regime of electro-optic sampling. This regime provides a solution for the limitations of the conventional perturbative regime. In contrast to the perturbative regime, the highly nonlinear nature of strong-field processes enables the generation of a continuum-like byproduct electric field, which can be used to measure transient electric fields across an exceptionally wide spectral range.

We developed a semiclassical model and benchmarked it against all our experiments with sub-3 fs optical pulses of octave-spanning bandwidth. Both experimentally and theoretically we observe a clear transition from a narrowband response in the perturbative regime to an extremely broadband response in the strong-field regime. We confirm that this broadband response arises from highly nonlinear processes, such as cascaded nonlinearities, Brunel radiation, and injection current. After benchmarking our model against all our measurements, we evaluate the full response and validate it against the first-principles *ab initio* TDDFT calculation. We have also developed a software (see Data availability section) for the evaluation of the response function for specific experimental conditions.

Our work paves the way towards realization of broadband fieldoscopic studies with simple electro-optic sampling scheme. In future experiments, it could be applied for time-domain measurements of extremely fast sub-cycle processes. The all-optical nature of this approach could be further utilized in imaging applications for the full spatio-temporal electric field information.

## 5. Materials and methods

### Semiclassical model

Electro-optic sampling signal $S$ originates from the heterodyne interference of LO and the BPF at the detection frequency $\omega_D$ (Eq. 2 of the main text). In the experiment the detection frequency $\omega_D$ is chosen by placing a narrowband bandpass filter after the non-linear crystal. In the model the same bandpass filter is applied numerically.

Since the byproduct field originates from a combined interaction of sampling and test fields at all points within a medium, it is necessary to model electric fields of the sampling, test and the byproduct fields at each point and time within the medium. For this we numerically solve the unidirectional pulse propagation equation without slowly varying envelope approximation:

$$\frac{\partial E_{Inc}(z,\omega)}{\partial z} = -i\left(\frac{\left[\sqrt{\epsilon(\omega)} - n_g\right]\omega}{c} - \beta(\omega_0)\right)E_{Inc}(z,\omega) - \frac{i\omega}{2c\sqrt{\epsilon(\omega)}}P_{\mathrm{NL}}(z,\omega), \tag{1}$$

where $E_{Inc}(z,\omega) = \hat{F}E_{Inc}(z,t) = \int_{-\infty}^{\infty} E_{Inc}(z,t)\exp(-i\omega t)\,dt$ is the Fourier transform $\hat{F}$ of the incident electric field $E_{Inc}(z,t)$, $z$ is the propagation coordinate, $\epsilon(\omega)$ is the linear dielectric permittivity, $n_g$ is the group refractive index, $\beta(\omega) = \sqrt{\epsilon(\omega)}\omega/c$, and $P_{\mathrm{NL}}(z,\omega)$ is the Fourier transform of the nonlinear part of the polarization.

In the time domain, the nonlinear term in the above expression can be written as

$$\frac{\partial E_{Inc}(z,t)}{\partial z} = -\frac{1}{2cn_0\epsilon_0}\frac{\partial P_{NL}(z,t)}{\partial \mathrm{t}}, \tag{2}$$

whereby, we have approximated $\epsilon(\omega)$ by the literature value of the refractive index $n_0$ of a medium for a characteristic frequency of the sampling pulse.

This equation is a forward-propagation equation, i.e. the backward-propagating wave in the sample is neglected. This assumption is justified as long as the nonlinear modification of the dielectric function, e.g. by the plasma contribution, is small. This condition holds as long as the metallization of the sample by the free carriers (which is usually accompanied by damage) is not reached. We also neglect the transverse dynamics (self-focusing, diffraction, etc.), since the beam is sufficiently wide (several tens of micrometers), and the propagation length is no longer than 12 μm (thickness of the sample used in the experiment). The slowly varying envelope approximation is not used, and $E_S(z,t)$ represents the real field and includes the carrier oscillations. This approach yields a unified treatment of an incident field with arbitrary spectral content, which is critical for extremely broad spectra considered in this work. The dielectric function $\epsilon(\omega)$ was calculated using the standard Sellmeyer equation and includes losses for the frequencies above the bandgap of quartz. The group refractive index $n_g$ was chosen so that the pulse does not shift in the time domain during the propagation.

In our model for non-linear polarizations, we include second- (second harmonic generation, optical rectification etc.) and third-order (self-phase modulation, cross-phase modulation, four-wave mixing, third harmonic generation, self-steepening, etc.) perturbative light-matter interaction as well as the strong-field effects described by the injection current, Brunel dynamics and cascaded non-linearities.

$$P_{NL}(z,t) = P_{\chi^{(2)}}(z,t) + P_{\chi^{(3)}}(z,t) + P_{Injection}(z,t) + P_{Brunel}(z,t). \tag{3}$$

We neglect higher-order perturbative nonlinearities ( $\chi^{(5)}$ , $\chi^{(7)}$ , etc.). The reliable characterization of these nonlinearities is not currently available, and in many cases, they are associated with photoionization, which is included in our model. We also do not consider harmonics generated by the three-step mechanism in the solid state (photoionization to the conduction zone, acceleration of almost-free carriers, followed by the recombination), since it is usually characterized by the emission of high-energy photons with energies above bandgap, which are not relevant for the present study.

*Propagation of the sampling pulse*

The second- and third order non-linear polarizations from the sampling field $E_S(z,t)$ are expressed in terms which are standard to perturbative non-linear optics:

$$P_{\chi^{(2)}}(z,t) = \epsilon_0 E_S \chi^{(2)}(z,t)^2. \tag{4}$$

$$P_{\chi^{(3)}}(z,t) = \epsilon_0 E_S \chi^{(3)}(z,t)^3. \tag{5}$$

The injection current and Brunel dynamics contributions are determined by the average position $\langle d\rangle(z,t)$ of the electron with respect to the equilibrium position in the parent "molecule". Here, by a "molecule", we denote an atom or a group of atoms of the solid-state which provides a single ionization event. Such treatment is itself an approximation, which is justified by the large bandgap of quartz and the corresponding strong localization of the electrons. Furthermore, it is determined by the relative ionization $\rho(z,t)$, which is the ratio of the conduction-band electron density to the density of "molecules":

$$P_{Injection+Brunel}(z,\omega) = -N_{\mathrm{mol}} e \hat{F}[\langle d\rangle(z,t)\rho(z,t)], \tag{6}$$

here $N_{\mathrm{mol}}$ is the concentration of the molecules and $e = 1.6 \times 10^{-19}$ is the electron charge. The dynamics of the $\langle d\rangle(z,t)\rho(z,t)$ is given by

$$\frac{\partial\big(\langle d\rangle(z,t)\rho(z,t)\big)}{\partial t} = \langle v\rangle(z,t) + x_0\Gamma(t), \tag{7}$$

where $\langle v\rangle$ is the average electron velocity and $x_0(z,t) \simeq -I_p/eE_S(z,t)$ is the initial displacement of the electron after the ionization event [41], where $I_p$ is the bandgap of a solid medium. We describe the ionization rate $\Gamma(t)$ by Ivanov-Yudin formalism [49]. Here we use the expression for $x_0$ provided by the tunneling exit in an isolated atom, which is again justified by the large bandgap.

The second term on the right-hand side of Eq. (7) describes the injection current. If expressed in time domain, it results in the following term in the propagation equation:

$$-\frac{1}{2c\epsilon_0}\frac{\partial P_{Injection}(z,t)}{\partial \mathrm{t}} = -\frac{N_{\mathrm{mol}} I_p \Gamma[E_S(z,t)]}{2c\epsilon_0 n_0 E_S(z,t)}, \tag{8}$$

The first term of Eq. (7) describes the Brunel radiation and is determined by the time evolution of $\langle v\rangle(z,t)\rho(z,t)$, as given by the second Newton's law in the form

$$\frac{\partial\big(\langle v\rangle(z,t)\rho(z,t)\big)}{\partial t} = -\frac{eE_S(z,t)}{m_e}\rho - \nu\langle v\rangle(z,t)\rho(z,t), \tag{9}$$

where $m_e$ is the effective electron mass and $\nu$ is the electron collision rate. Here we disregard the initial velocity of the electron after the ionization, as well as the anisotropy of the effective mass as well as its dependence on the position in the Brillouin zone. If expressed in time domain and for $\nu = 0$, it results in the following simple term in the propagation equation:

$$-\frac{1}{2c\epsilon_0}\frac{\partial P_{Brunel}(z,t)}{\partial t} = -\frac{N_{mol}e^2}{2c\epsilon_0 m_e}\int_{-\infty}^{t} E_S(z,t')\rho(z,t')dt', \tag{10}$$

Finally, we evaluate $P_{\chi^{(2)}}(z,t)$, $P_{\chi^{(3)}}(z,t)$, $P_{Injection}(z,t)$, $P_{Brunel}(z,t)$ using Eqs. 4, 5, 8, 10 and substitute the results into the Eq. 3 to obtain a total polarization nonlinear polarization $P_{NL}(z,t)$ at each point within the medium (z) and each time stamp (t), which we further substitute into the Eq. 1 to obtain the overall electric field $E_{S'}(z,t)$ of the propagated sampling pulse.

_Propagation of the test pulse_

In the experiment as well as in the model, we set the electric field of the test pulse to be weak such that by itself its field strength is not enough for substantial non-linear light-matter interaction. Therefore $P_{\text{NL}}(z,\omega) \approx 0$. Similarly to the sampling pulse, the propagated test electric field $E_{T'}(z,t)$ is then obtained using Eq. 1.

_Propagation of the byproduct electric field_

The byproduct field originates from the combined light-matter interaction of the propagated electric field of the sampling pulse $E_{S'}(z,t)$ and the propagated electric field of the test $E_{T'}(z,t)$ pulse. As we describe in the main text, in our study sampling and test pulses are orthogonally polarized with respect to each other. The symmetry (z-cut α-quartz) and the orientation of the crystal was chosen such that the polarization of $\chi^{(2)}$ byproduct fields were the same as the polarization of the sampling field. The polarization $\chi^{(3)}$ of byproduct fields on the other hand was the same as the polarization of the test field. In the experiment we measured the $S(\tau)$ signal in the polarization of the test field only, therefore the $\chi^{(2)}$ byproduct fields are filtered out, resulting $P_{\chi^{(2)}}(z,t) = 0$. The third-order byproduct non-linear polarization between the sampling $E_{S'}(z,t-\tau)$ and test $E_{T'}(z,t)$ fields are expressed as:

$$P_{\chi^{(3)}}(z,t) = \epsilon_0 \chi^{(3)} E_{S'}(z,t-\tau)(z,t)^2 E_{T'}(z,t), \tag{11}$$

where $\tau$ is the time delay between sampling and test pulses.

The byproduct non-linear polarization for the injection current is similar to Eq. 8, however expressed in terms of the combined electric field:

$$-\frac{1}{2c\epsilon_0}\frac{\partial P_{Injection}(z,t)}{\partial \mathrm{t}} = \frac{N_{\text{mol}} I_p \Gamma[|E_{S'}(z,t-\tau) + E_{T'}(z,t)|]}{2c\epsilon_0 n_0 (|E_{S'}(z,t-\tau) + E_{T'}(z,t)|)^2}(E_{S'}(z,t-\tau)\vec{e_S} + E_{T'}(z,t)\vec{e_T}), \tag{12}$$

where $\vec{e_S}$ and $\vec{e_T}$ stand for unit vectors of the polarizations of the sampling and test fields. This equation consists of an ionization-related factor multiplied by the electric field. The ionization-related factor however, is determined by the ionization rate which depend on the absolute value of the total electric field $|E_{S'}(z,t-\tau) + E_{T'}(z,t)| = \sqrt{E_{S'}^2(z,t-\tau) + E_{T'}^2(z,t)}$. For weak test pulses with a polarization orthogonal to the sampling pulse, the absolute value determined predominantly by the sampling pulse, with test pulse giving only a negligibly small correction. On the other hand, since the measurement is conducted along the $\vec{e_T}$ polarization, the orthogonal $E_{S'}(z,t-\tau)\vec{e_S}$ term in the $\vec{e_T}$ polarization is absent. With this, the expression for the byproduct non-linear polarization due to the injection current takes the following form:

$$-\frac{1}{2c\epsilon_0}\frac{\partial P_{Injection}(z,t)}{\partial \mathrm{t}} = \frac{N_{\text{mol}} I_p \Gamma[|E_{S'}(z,t-\tau)|]}{2c\epsilon_0 n_0 (|E_{S'}(z,t-\tau)|)^2} E_{T'}(z,t). \tag{13}$$

Applying similar considerations, the expression for the byproduct non-linear polarization due to the Brunel dynamics can be written as:

$$-\frac{1}{2c\epsilon_0}\frac{\partial P_{Brunel}(z,t)}{\partial t} = -\frac{N_{mol}e^2}{2c\epsilon_0 m_e}\int_{-\infty}^{t} E_{T'}(z,t')\rho(z,t')dt', \tag{14}$$

Similarly, as in the case of the sampling pulse propagation, we evaluate $P_{\chi^{(3)}}(z,t)$, $P_{Injection}(z,t)$, $P_{Brunel}(z,t)$ using Eqs. 11, 13, 14 and substitute the results into the Eq. 3 to obtain a total nonlinear polarization $P_{NL}(z,t)$ at each point within the medium ($z$) and each time stamp ($t$), which we further substitute into the Eq. 1 to obtain the overall byproduct electric field $E_{BPF}(z,t,\tau)$ for various delays between incident sampling and test pulses.

*Signal and the response function*

To compute the signal measured in the experiment $S(\tau)$, we take the resulting $E_{BPF}(t,\tau)$ at the end of the medium. In general, the local oscillator is the overall $E_{S'}(t,\tau)$ electric field at the end of the medium ($E_{LO}(t,\tau) = E_{S'}(t,\tau)$). However, since the measurement of the $S(\tau)$ is conducted along the polarization of the test pulse, only the component of the $E_{S'}(t,\tau)$ in the polarization of the test pulse (second harmonic field) constitutes the local oscillator. We take $E_{BPF}(t,\tau)$ and $E_{LO}(t,\tau)$ at the experimental detection bandwidth (see section S12 in the Supplementary Material) and substitute into the Eq. 2 to obtain $S(\tau)$ which can be compared with the experiment.

To obtain the response function $R(\omega)$, that can be used for sampling of any arbitrary input test electric field (Eq. 3), we first note that in certain cases (more details in the Supplementary Material), this function can be approximated by a simple analytic expression. In a general case however, the response function is obtained by evaluating $S(\tau)$ numerically for a known sampling pulse and a test pulse represented by a delta function. In this case $R(\omega) = S(\omega)$.

**Ab-initio TDDFT model**

To validate the results obtained from the semiclassical model, we have conducted an *ab initio* first-principle time-dependent density-function theory (TDDFT) simulation which is based on solving time-dependent Kohn-Sham equation in real time and real space with norm-conserving pseudopotentials.

**Experimental methods**

The laser beamline used in the experiments (see Supplementary Material for details) consists of a Ti:sapphire oscillator (Rainbow 2, Spectra Physics), followed by chirped-pulse amplification to a pulse energy of 1 mJ at a 3 kHz repetition rate. The amplified pulses undergo spectral broadening in a hollow-core fiber and are subsequently compressed using a chirped mirror compressor. Experimental data acquisition (details in the Supplementary Material) was performed using a dual-phase lock-in amplifier (SR-830, Stanford Research Systems) and a grating spectrometer (Ocean Optics).

**Funding**

This study is funded by the Swiss National Science Foundation, grants no. TMPFP2 217068 and PZ00-2_232911.

## Acknowledgments

The authors thank Ferenc Krausz and Max Planck Institute of Quantum Optics for the experimental infrastructure.

## Author contributions

## Disclosures

Authors declare that they have no competing interests.

## Data availability

The datasets generated during and/or analyzed during the current study are available from the corresponding author upon reasonable request.

The software for the evaluation of the experimental response function can be downloaded from the opensource GitHub repository: https://github.com/AHusakou/SF-EOS

## Supplemental document

See Supplement_Material for supporting content.